\documentclass[twocolumn,showpacs,preprintnumbers,amsmath,amssymb,prd,floatfix]{revtex4}
\usepackage{rotating}
\usepackage{graphicx}
\usepackage{dcolumn}
\usepackage{bm}
\usepackage{multirow}

\begin{document}

\title{First ground based measurement of atmospheric Cherenkov light from cosmic rays} 

\author{F. Aharonian$^{1}$}
\author{A.G.~Akhperjanian $^{2}$}
\author{A.R.~Bazer-Bachi $^{3}$}
\author{M.~Beilicke $^{4}$}
\author{W.~Benbow $^{1}$}
\author{D.~Berge $^{1}$} \thanks{now at CERN, Geneva, Switzerland}
\author{K.~Bernl\"ohr $^{1,5}$}
\author{C.~Boisson $^{6}$}
\author{O.~Bolz $^{1}$}
\author{V.~Borrel $^{3}$}
% ...

\affiliation{$^{1}$
\footnotesize
Max-Planck-Institut f\"ur Kernphysik, P.O. Box 103980, D 69029
Heidelberg, Germany
}\affiliation{$^{2}$
 Yerevan Physics Institute, 2 Alikhanian Brothers St., 375036 Yerevan,
Armenia
}\affiliation{$^{3}$
Centre d'Etude Spatiale des Rayonnements, CNRS/UPS, 9 av. du Colonel Roche, BP
4346, F-31029 Toulouse Cedex 4, France
}\affiliation{$^{4}$
Universit\"at Hamburg, Institut f\"ur Experimentalphysik, Luruper Chaussee
149, D 22761 Hamburg, Germany
}\affiliation{$^{5}$
Institut f\"ur Physik, Humboldt-Universit\"at zu Berlin, Newtonstr. 15,
D 12489 Berlin, Germany
}\affiliation{$^{6}$
LUTH, UMR 8102 du CNRS, Observatoire de Paris, Section de Meudon, F-92195 Meudon Cedex,
France
}
% ...

\begin{abstract}
A recently proposed novel  technique for the detection of cosmic rays with arrays of \emph{Imaging Atmospheric  Cherenkov Telescopes} is applied to data from the \emph{High Energy Stereoscopic System} (H.E.S.S.). The method relies on the ground based detection of Cherenkov light emitted from the primary particle prior to its first interaction in the atmosphere. The charge of the primary particle (Z) can be estimated from the intensity of this light, since it is proportional to Z$^2$. Using H.E.S.S. data, an energy spectrum for cosmic-ray iron nuclei in the energy range 13--200 TeV is derived. The reconstructed spectrum is consistent with previous direct measurements and is one of the most precise so far in this energy range. 
\end{abstract}

\maketitle

\section{Introduction}

\subsection{Motivation}

Cosmic rays reach the earth at a rate of approximately 1000 s$^{-1}$ m$^{-2}$. Their energy spectrum is steeply falling and remarkably featureless over ten orders of magnitude in energy \cite{HOERANDEL}.  The differential flux is well described by a power law ($\Phi \sim E^{-\gamma}$) with a steepening of the spectrum at a few PeV (the so-called ``knee'') and a flattening around 10 EeV (the so-called ``ankle''). Despite advances in the field, the origin of cosmic rays is still unresolved. Supernova explosions are thought to be the major contributor  at energies up to 1 PeV \cite{SNR,MHILLAS}, but conclusive proof is still missing. The High Energy Stereoscopic System (H.E.S.S.) \cite{HESS} has clearly identified supernova shock waves as sources of high-energy particles \cite{RXJ, VELAJR}. However, the nature of these particles -- electrons or cosmic-ray nucleons -- remains under debate. 

The elemental composition of cosmic rays is similar to the composition of the solar system, if one accounts for propagation effects through the galaxy \cite{LONGAIR}. At present the best measurements of elemental spectra in the energy range 1 GeV to 0.5 PeV come from long duration balloon flights \cite{RUNJOB}. Because of the decreasing flux of cosmic rays and the limited collection area of these experiments ($\approx$1 m$^2$), it is hard to extend such measurements to higher energies. A further improvement in the accuracy and energy range of composition measurements could provide crucial information about the acceleration mechanism and propagation of these particles, and therefore provide further clues about their origin.

In 2001, Kieda et al. \cite{KIEDA} proposed a new method for the measurement of cosmic rays with \emph{\textbf{I}maging \textbf{A}tmospheric \textbf{C}herenkov \textbf{T}elescopes} (IACTs). The central idea of this method is to detect the Cherenkov light emitted by \emph{primary} cosmic-ray particles (so-called \emph{\textbf{D}irect \textbf{C}herenkov light}) from the ground. While DC-light has been measured in the past by balloon experiments \cite{SOOD,CLEM}, the measurement from the ground takes advantage of the huge detection area ($\approx$$10^5$ m$^2$) of IACTs, in principle enabling the extension of spectral and composition measurements up to $\sim$ 1 PeV. Here we review this technique and describe its application to data from H.E.S.S. We present the measurement of the iron spectrum and give an outlook on future applications of this method.

\subsection{Technique}

\begin{figure}
  \begin{center}
    \includegraphics[width=9cm]{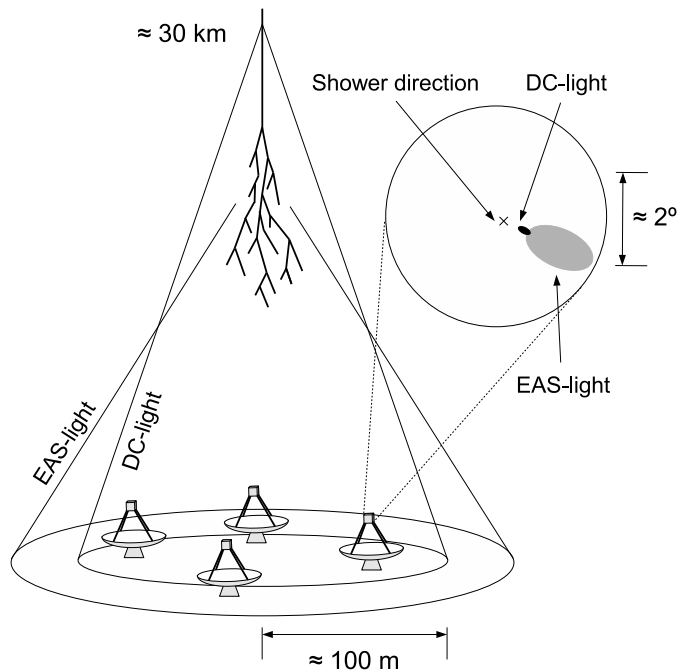}
  \end{center}
  \caption{Schematic representation of the Cherenkov emission from a cosmic-ray primary particle and the light distribution on the ground and in the camera plane of an IACT.}
  \label{sketch}
\end{figure}

When cosmic rays enter the atmosphere they emit Cherenkov light above an element-dependent energy threshold. The Cherenkov angle increaces with the density of the surrounding medium. The emission angle of the DC-light therefore increases with increasing depth of the primary particle in the atmosphere, creating a light cone on the ground with a radius of roughly 100 m (see Fig. \ref{sketch}).  At a typical height of 30 km the particle interacts and a particle cascade is induced (Extensive Air Shower, EAS). The Cherenkov light from these secondary particles creates a second, wider, light cone on the ground.

\begin{figure}
  \begin{center}
    \includegraphics[width=9cm]{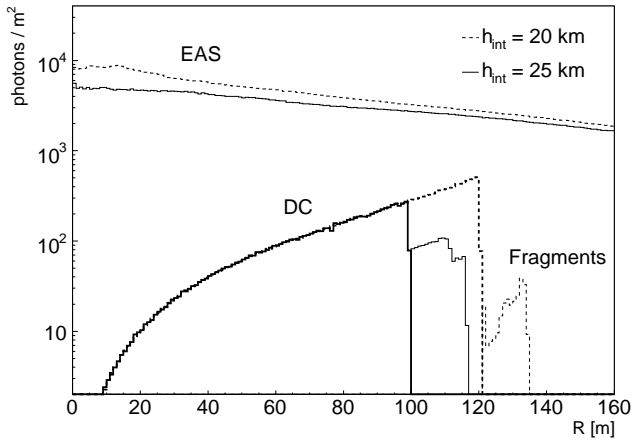}
  \end{center}
  \caption{Simulated intensity distribution on the ground for the EAS-light and DC-light of an individual 50 TeV iron nucleus, as a function of distance from the shower core, for two different first interaction heights (the shower core is defined as the intersection point of the shower axis on the ground). The zenith angle is $0^\circ$. The drop in DC-intensity at 100/120 m reflects the first interaction height.  The low intensity tail at larger radii is caused by Cherenkov light from fragments of the primary nucleus.}
  \label{intground}
\end{figure}

The intensity of the DC-light is proportional to the square of the charge Z of the emitting particle, and can therefore be used to identify the primary particle. The challenge for detecting DC-light is to distinguish it from the much brighter EAS-light background (Fig. \ref{intground}). Because the DC-light is emitted higher in the atmosphere, it is emitted at a smaller angle than the EAS-light, and is therefore imaged closer to the shower direction in the camera plane. A typical  emission angle for DC-light is $0.15^\circ$ to $0.3^\circ$, whereas most of the EAS-light is emitted at angles greater $0.4^\circ$ from the direction of the primary particle (for a more detailed discussion see \cite{KIEDA}). Cherenkov cameras, with pixel sizes of $\sim 0.1^\circ$ are therefore able to resolve the DC-emission as a single bright pixel between the reconstructed shower direction  and the \emph{center of gravity} (cog) of the EAS-image in the camera plane (Fig. \ref{sketch}). 

\begin{figure}
  \begin{center}
    \includegraphics[width=9cm]{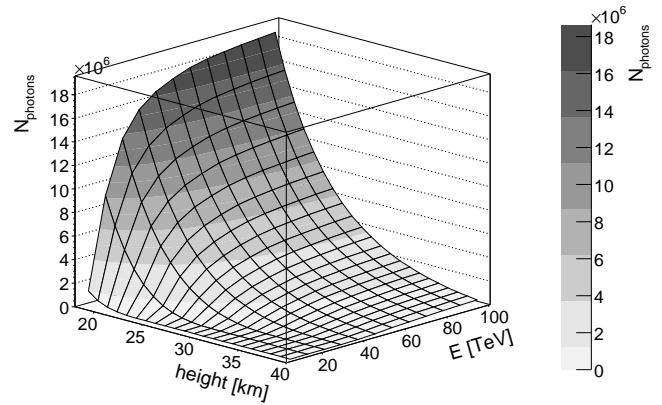}
  \end{center}
  \caption{Total number of emitted DC-photons as a function of energy and first interaction height for an iron nucleus at a zenith angle of $0^{\circ}$, calculated using an atmospheric profile appropriate for the H.E.S.S. site.}
  \label{nphot}
\end{figure}

Unfortunately, the number of emitted DC-photons also depends on the emission height and on the energy of the primary particle (Fig. \ref{nphot}). The height of first interaction of hadrons typically varies between 20 to 40 km, hence the total amount of emitted DC-photons of particles with the same atomic number and energy varies significantly.   As will be discussed later, this makes the estimation of the primary charge more difficult.

The energy range to which this technique can be applied depends on the charge of the primary particle \cite{KIEDA}. At lower energies the limiting factor is that the primary particle momentum must exceed the Cherenkov threshold. At very high energies, the EAS-light outshines the DC-light, making the detection of the latter impossible. The reason for this is that the intensity of the EAS-light increases approximately linearly with energy, whereas the amount of emitted DC-photons remains basically constant above a certain energy (see Fig. \ref{nphot}).

The aim of the current work is to demonstrate that the technique of DC-light detection can be applied to instruments such as H.E.S.S. to measure the flux of cosmic-ray iron nuclei. Because of their large atomic number and high flux compared to other heavy elements, iron nuclei are well suited for DC-light detection. The lower energy threshold for the detection of these nuclei is $\sim 10$ TeV. 

The H.E.S.S. instrument used for this measurement consists of four IACTs situated in the Khomas highland of Namibia, at a height of 1830 m above see level. Each telescope is equipped with a 960 pixel camera. Each pixel has an angular diameter of $0.16^\circ$, providing a total field of view of 5$^\circ$  diameter.  The four telescopes are triggered in coincidence and image the Cherenkov light from EASs. The exact trigger conditions can be found in \cite{TRIGGER,CRAB} and are far below the applied analysis cuts described in the next section. As will also be described there, the properties of the primary particle (such as direction or energy) can be reconstructed from the shower images.

H.E.S.S. is a $\gamma$-ray experiment. The main challenge in detecting $\gamma$-rays is to distinguish them from the much larger background of hadronic cosmic rays (see for example \cite{CRAB}). This background, recorded during normal $\gamma$-ray observations, is now used to search for events with DC-light.

\section{Event selection and reconstruction}

\subsection{Shower reconstruction and candidate event selection}

As for the standard H.E.S.S. analysis, the raw shower images are calibrated \cite{CALIB} and the pixel intensities are corrected to account for the loss of optical efficiency of the system over time \cite{CRAB}. Afterwards the images are cleaned to remove low intensity substructure in the shower images and hence improve shower reconstruction. The image cleaning consists of a two-staged tail-cut, which requires pixels to have an intensity greater than 20 (10) photo-electrons (pe) and a neighboring pixel with an intensity of 10 (20) pe. Afterwards, an ``island cleaning'' is applied, where all pixels that are not connected to the pixel with the maximum intensity by neighboring pixels are removed. For a further reduction of shower fluctuations the same procedure is applied again with stronger cuts (200 and 100 pe). By default the soft-cleaned images are used for the shower reconstruction. The strong-cleaned images are only used instead if they contain more than 7 pixels in more than two camera images.

The showers are reconstructed using the standard \emph{stereoscopic Hillas analysis} \cite{HILLAS}, whereby the shower (and therefore particle) direction and  the intersection point of the shower axis on the ground are reconstructed by intersecting the major axes of the different shower images. The energy of the primary particles is reconstructed from the total image intensity $I_{\rm{tot}}$, the impact parameter $R_{\rm{core}}$ (perpendicular distance from the shower axis to the telescope) and the zenith angle $\theta$, by comparing these parameters to simulations. The mean EAS-light yield at a fixed energy varies with the atomic number of the primary particle, which introduces a systematic shift in the energy reconstruction between different elements. In this analysis all energies are reconstructed under the assumption that the primary particle is an iron nucleus. 

This energy reconstruction technique leads to a systematic bias close to the energy threshold for detection \cite{CRAB}. Therefore only events with a reconstructed energy greater than 13 TeV were considered, for which the energy bias is less than 5\% (the exact energy value is 12.59 TeV, which corresponds to $\log_{10}(E/\rm{TeV}) = 1.1$ and will always be referred to as 13 TeV in the following). Additionally, to avoid images truncated at the camera edge, only images with a center of gravity less than 2$^\circ$ from camera center are used. Finally, to select well reconstructed showers, only events that contain at least two camera images with an \emph{Aspect Ratio}  smaller than 0.75 (\emph{Aspect Ratio} $\equiv \frac{image~width}{image~length}$) were considered.

\subsection{DC-light detection}

\begin{figure}
  \begin{center}
    \includegraphics[width=9cm]{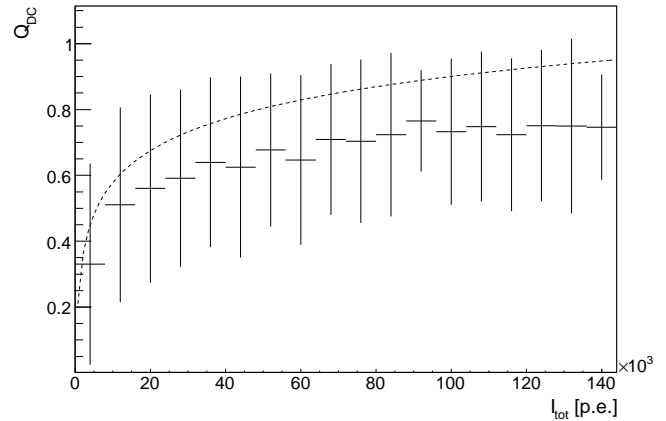}
  \end{center}
  \caption{Mean DC-ratio $Q_{\rm{DC}}$ as a function of the total intensity of the camera image $I_{\rm{tot}}$ for iron events from simulations (events which interacted in the atmosphere before passing the Cherenkov energy threshold were not considered since they contain no DC-light). The zenith angle is 0$^{\circ}$. The error bars show the RMS of the distribution in each bin. The dotted line shows the selection function given in table \ref{cuts}.}
  \label{qdccut}
\end{figure}

DC-Light can be identified as a single high intensity pixel between the reconstructed shower direction and the cog of the EAS-shower in the camera images (Fig. \ref{sketch}). The main selection parameter for finding this DC-pixel in the camera image is the \emph{DC-ratio}, defined as:
\begin{equation}
Q_{\rm{DC}}=\frac{I_{\rm{max. neighb.}}}{I_{\rm{pixel}}} ,
\end{equation}
where $I_{\rm{max. neighb.}}$ is the maximum intensity of the neighboring pixels. The pixel with the minimum $Q_{\rm{DC}}$ is determined in the relevant angular region in each camera image. The parameters used to constrain this region are the angular distance from the DC-pixel to the shower direction ($\Delta_{\rm{DC}}^{\rm{dir}}$), to the cog of the EAS ($\Delta_{\rm{DC}}^{\rm{cog}}$) and to the line connecting these two points ($\Delta_{\rm{DC}}^{\rm{\perp}}$). Afterwards a selection on the impact parameter is applied to ensure that the telescope is inside the DC-light cone. Finally, to avoid detector saturation effects, the intensity of the DC-light candidate pixel $I_{\rm{DC-pixel}}$  has to be below the saturation intensity. The exact cut values for the identification of a DC-pixel were optimized using iron simulations and are summarized in Table \ref{cuts}.
\begin{table}
\begin{tabular}{l|l}
\textbf{Parameter}	         & \textbf{Cut Condition}                           \\ \hline
$Q_{\rm{DC}} $   	         & $< 0.14 \ln(\frac{I_{\rm{tot}}[p.e.]/161}{\cos(\theta) })$  \\
$\Delta_{\rm{DC}}^{\rm{dir}} $   & $< 0.45^\circ$	                            \\ 
$\Delta_{\rm{DC}}^{\rm{cog}}$    & $> 0.17^\circ$	                            \\
                                 & $< 0.91^\circ$	                            \\ 
$\Delta_{\rm{DC}}^{\rm{\perp}}$  & $< 0.23^\circ$                                   \\ 
$R_{\rm{core}}$   	         & $>$ 40 m                                         \\
                                 & $<$ 170 m                                        \\ 
$I_{\rm{DC-pixel}} $   	         & $< 2500$ pe
\end{tabular}
\caption{Cut parameters for DC-pixel detection. For definitions of the parameters see text. }
\label{cuts}
\end{table}
To illustrate the applied cuts on one example, the distribution of $Q_{\rm{DC}}$ for iron nuclei and the cut values are shown in Fig. \ref{qdccut} as a function of the total image intensity $I_{\rm{tot}}$. The mean $Q_{\rm{DC}}$ value depends on $I_{\rm{tot}}$ because at higher energies more and more Cherenkov light from secondary particles falls into the angular region of the DC-light, while the DC-intensity remains basically constant above a certain energy threshold. 

Once a pixel in the camera image fulfills all the mentioned selection parameters the DC-light intensity $I_{\rm{DC}}$ is reconstructed by subtracting the mean intensity of the neighboring pixels $I_{\rm{neighb. pixels}}$ from the DC-pixel intensity:
\begin{equation}
I_{\rm{DC}}=I_{\rm{DC-pixel}} - <I_{\rm{neighb. pixels}}>
\end{equation}

\section{Data analysis and Simulations}

\subsection{Simulations}

\begin{table}
\begin{tabular}{l|c|c}
  Z range 	&  representative element & $F_{13}^{200}$ [$10^{-4}$ s$^{-1}$sr$^{-1}$ m$^{-2}$]	\\ \hline
1-6		& p 	(Z=1)		&12.6	\\
7-9		& O 	(Z=8)		&1.43	\\
10-16		& Mg 	(Z=12)		&2.09	\\
17-24		& Ca 	(Z=20)		&0.56	\\
25-28		& Fe	(Z=26)		&2.50	
\end{tabular}
\caption{Representative element and integrated flux between 13 and 200 TeV, $F_{13}^{200}$, of the reference composition \cite{HOERANDEL,WIEBEL} for the five charge bands of the simulated flux. }
\label{zbans}
\end{table}

As for all air shower experiments, the present analysis relies on comparisons to Monte Carlo simulations. Simulations are used to calibrate  the energy and charge (section \ref{sec:charge}) estimation and to determine the detection efficiency of the system (section \ref{sec:flux}). In order to perform a direct comparison between simulations and data, simulations have been produced for five different elements, representative of five different charge bands (see Table \ref{zbans}). The charge bands cover a $Z$ range from 1 to 28. The contribution of elements with $Z>28$ is likely negligible: although no flux measurements for these \emph{ultra heavy cosmic rays} exist in the energy region of importance here, measurements of energy spectra for elements with $Z>28$ at lower energies \cite{CRIS,HOERANDEL} and flux predictions \cite{UHCR} in the TeV energy region are more than three orders of magnitude below the flux of the iron band.

The contribution of each charge band to the total simulated flux and the energy distribution of the events inside each band are weighted according to the measured fluxes given in \cite{WIEBEL}  and \cite{HOERANDEL} (referred to in the following as \emph{reference composition}). Ref. \cite{WIEBEL} shows a parameterization of the different elemental spectra, obtained by combining measurements from several experiments. The errors given on the absolute flux normalization are $\leq$10\%. However, they probably do not reflect the entire uncertainty in the spectra, since systematic uncertainties may have been underestimated in the individual data sets \cite{WIEBEL}.  Additionally, in the energy range of interest here, specifically for the calcium band, the values given are extrapolated from measurements at lower energies. Ref. \cite{HOERANDEL} presents the same parameterization with more recent data for the proton, helium and iron flux. The normalization of the flux for these elements differs by approximately 25\% between the two parameterizations. For the mentioned reasons we assume this difference to be a realistic error in the integral fluxes of the different charge bands between 13 and 200 TeV.  In the following comparisons between the data and the simulations, this error is always included. Since the 25\% error is still somewhat arbitrary, we will also discuss the effect of a more conservative error of 50\% on the presented measurements at the end of this work.

The shower parameters of relevance here depend on the details of high energy hadronic interactions. To assess the systematic errors arising from uncertainties in these interactions, the analysis is performed with simulations based on two independent hadronic interaction models, SIBYLL 2.1 \cite{SIBYLL} and QGSJET 01f \cite{QGSJET}. (The newer version QGSJET 02 \cite{QGSJETII} was not available at the time of the analysis. However, the hadronic interaction uncertainty estimation with QGSJET 01f should be more conservative, since more recent hadronic models are expected to model the interactions more accurately). The simulations for both models were performed using the shower simulation program  CORSIKA 6.0321  \cite{CORSIKA}. For each model a total of $\sim 10^{6}$  showers were simulated in an energy range from 1 to 200 TeV. The zenith angle of the simulations was chosen to match the mean zenith angle of the data set (13.6$^\circ$) described in the next section.

\subsection{Data}

\begin{figure*}
  \begin{center}
    \includegraphics[width=15cm]{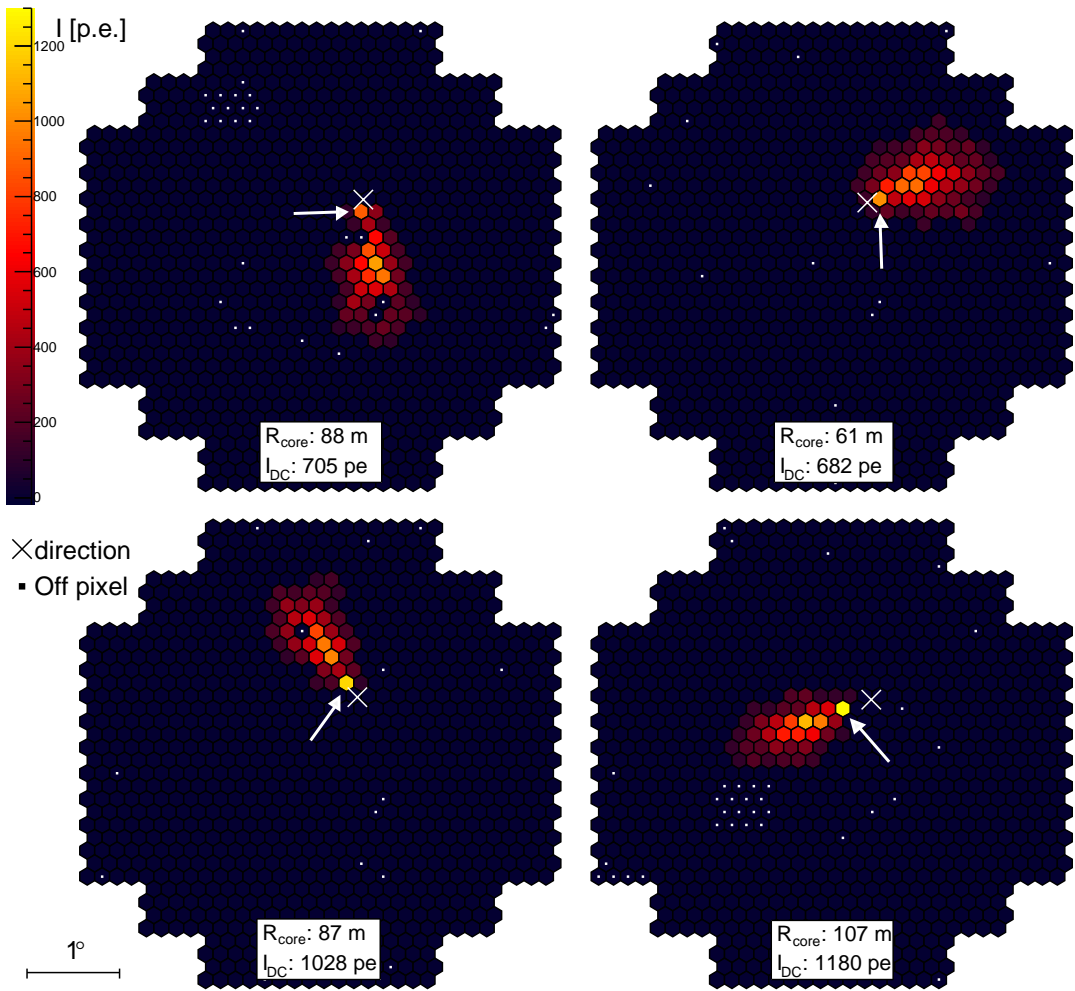}
  \end{center} 
  \caption{A measured event with indications of DC-light in all four cameras images (indicated by arrows), after high threshold image cleaning. The reconstructed shower direction is shown by a cross ($\times$) in each image. The reconstructed energy of this event is 50/48 TeV based on QGSJET/SIBYLL simulations. The reconstructed impact parameter and DC-light intensity for each telescope are shown in the lower panels in each image. The energy and impact parameter resolutions are $\approx$20\% and $\approx$20 m, respectively. The white points mark disabled pixels.}
  \label{figcam}
\end{figure*}

The data considered here were taken between 2004 and 2006 with the full four telescope H.E.S.S. array. Because cosmic rays are deflected by magnetic fields in the galaxy, their flux in the measured energy region is expected to be very close to isotropic \cite{LONGAIR}. Therefore it is possible to use all available H.E.S.S data, independent of the target position. However, to reduce systematic uncertainties due to zenith angle  effects on detection efficiencies and energy reconstruction, only data runs with a mean zenith angle smaller than 22$^\circ$ are considered. After standard quality selection criteria and dead time  correction, the data set amounts to 357 hours of observation time.

\begin{figure}
  \begin{center}
    \includegraphics[width=9cm]{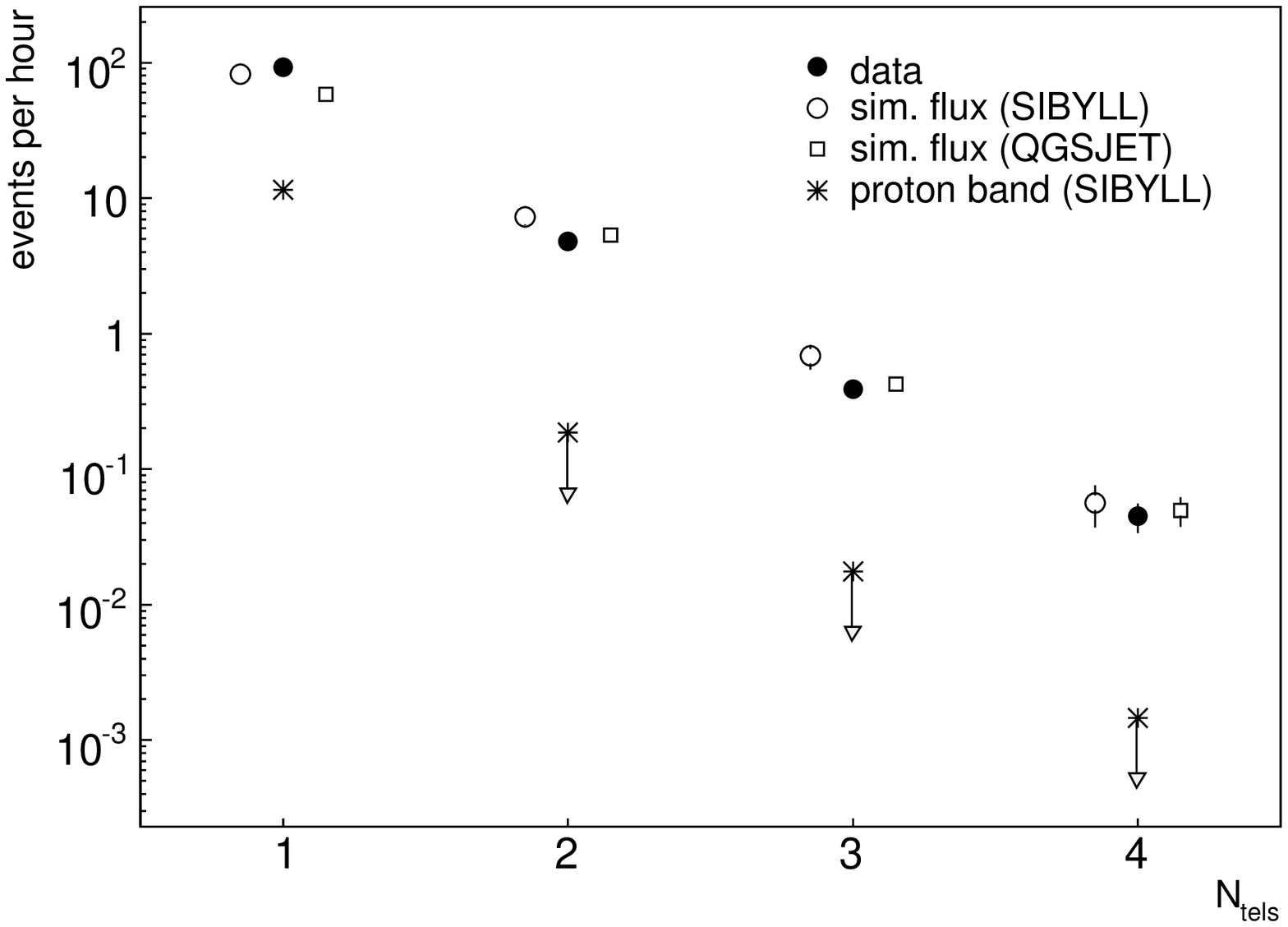}
  \end{center}
  \caption{Event rate for different telescope multiplicities $N_{\rm{tel}}$ compared to the expected rate from the simulation of the cosmic-ray flux using the SIBYLL and QGSJET interaction models. Additionally, the  background event rate from the proton band, estimated using SIBYLL simulations, is shown.  No background events were identified for $N_{\rm{tel}}\geq2$ and upper limits are shown (this result is the same for both hadronic models, which is why only the SIBYLL simulations are shown). An upper limit at 95\% confidence level is shown for $N_{\rm{tel}}=2$. The upper limits for $N_{\rm{tel}}>2$ are derived under the assumption that they have the same ratio to the detected event rate as the $N_{\rm{tel}}=2$ events. This is a conservative estimate since the fraction of misidentification is expected to decrease with increasing $N_{\rm{tel}}$. }
  \label{figntel}
\end{figure} 

In total, 35364 events in the energy region from 13 to 200 TeV passed the selection criteria. One example of an event with DC-light in all four telescopes is shown in Fig. \ref{figcam}. High intensity pixels close to the reconstructed shower direction are evident in all four images (indicated by arrows).  In agreement with the expected trend from Fig. \ref{intground}, the highest/lowest DC-pixel intensity also corresponds to the largest/smallest impact parameter, respectively. 

Fig. \ref{figntel} shows the DC-light detection rate as a function of the telescope multiplicity $N_{\rm{tel}}$ (number of telescopes in which DC-light is detected simultaneously) for data and simulations for both hadronic models. While the shape of the distributions agrees well, the event rate is higher by $\approx$25\% for the SIBYLL simulations. This difference between the models gives an estimate of the systematic error introduced in the analysis due to hadronic interaction uncertainties. Taking this systematic error into account, the simulated rates and the data are consistent.

\subsection{Background}

The detection rate of events with DC-light is expected to have some background due to misidentifications. These misidentifications can occur due to shower fluctuations, which can lead to single high intensity pixels in the EAS-light images. The rate of false detections can be estimated using proton simulations. Protons dominate the cosmic-ray flux in the energy region of interest (see Table \ref{zbans}). Additionally, protons emit only  a negligible  amount of DC-light compared to their EAS-light yield at these energies \cite{KIEDA}, therefore any detection of DC-light in proton simulations can be considered a fluctuation of the EAS-light yield and therefore a false detection.  As shown in Fig. \ref{figntel}, the expected misidentification rate for the proton band is $\approx$10\% of the measured event rate in the data for  $N_{\rm{tel}}$=1. However, for $N_{\rm{tel}}\geq$2 there are no misidentified events found in the proton band for either hadronic interaction model. The upper limit derived for the misidentification rate for these events is almost two orders of magnitude below the detected event rate in the data. Events with $N_{\rm{tel}}\geq$2 can therefore be considered as essentially background-free. 

In order to minimize systematic uncertainties due to background estimation in the presented analysis, only events with $N_{\rm{tel}}\geq$2 were considered. They will be refered to in the following as \emph{DC-events}. In total, 1899 DC-events were found in the data. The resolution of the shower parameter reconstruction for these events is $\approx$0.1$^\circ$ for the shower direction, $\approx$20 m on the shower core position and $\approx$15\% on the primary energy. 

\subsection{Primary charge reconstruction}
\label{sec:charge}

\begin{figure}
  \begin{center}
    \includegraphics[width=8cm]{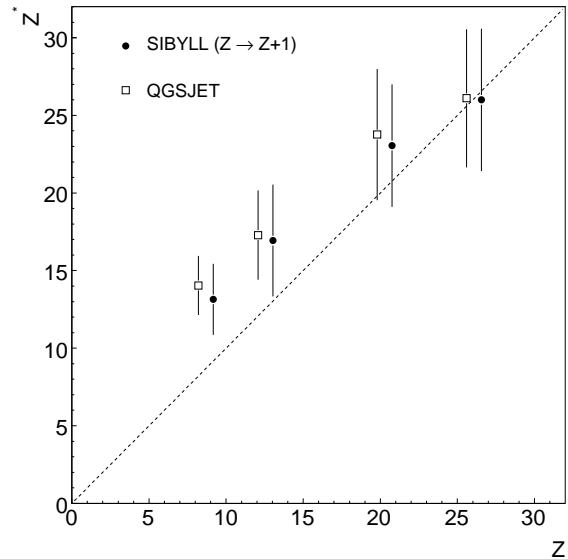}
  \end{center}
  \caption{Mean reconstructed charge $Z^*$ as a function of the true charge $Z$ for DC-events in an energy range of $1.5 < \log_{10}(E/\rm{TeV}) < 1.7$ for  both hadronic models. For clarity the x-axis is shifted by $+1$ for SIBYLL. The error bars show the RMS of the distribution in each bin.}
  \label{zoz}
\end{figure}

The elemental composition of the DC-events can be estimated using the $Z$ dependence of the DC-light intensity. The reconstructed charge $Z^{*}$ is defined as:   
\begin{equation}
Z^* = d(E,\theta) \sqrt{I_{\rm{DC}}} ,
\end{equation}
where $d(E,\theta)$ is a factor that normalizes the mean of the $Z^*$ distribution from iron simulations to the atomic number $Z$ of iron. The energy dependence of $d$ is due to the energy dependence on the number of emitted DC-photons, since the emission for iron nuclei is not saturated in the lower part of the observed energy range (Fig. \ref{nphot}). The zenith angle $\theta$ dependence of $d$  arises from the increasing distance between the average first interaction point to the telescopes with increasing $\theta$.

The charge resolution achieved using $Z^*$ is energy dependent and improves for higher energies, for two principal reasons:
\begin{enumerate}
\item The separation of the DC-light intensity distributions, and hence the $Z^*$ distributions, for different elements is maximized when the elements compared have high enough energies that their DC-light emission is saturated.  The saturation energy increases with charge. The heaviest element in this analysis is iron, for which the saturation energy is $\approx$50 TeV, which means that the charge separation of $Z^*$ continues to improve up to this energy.
\item A significant fraction of the detected DC-events are dominated by emission from secondary particles created in the first interaction. The reconstructed charge $Z^*$ for these events is lower than the charge of the primary particle. Simulations show that the fraction of DC-events dominated by the DC-light from secondary particles decreases with energy. For iron it is $\approx$60\% at low energies ($\approx$13 TeV) and drops to $\approx$35\% above the saturation energy.
\end{enumerate}

Fig. \ref{zoz}  shows the charge resolution obtained in an intermediate energy range for both hadronic models. The charge resolution achieved is not sufficient to assign the charge of the primary on an event-wise basis to one of the four charge bands. However, as will be shown later, it is possible to measure the fraction of elements belonging to the iron band in the data on a statistical basis, and therefore estimate the iron flux. 

The main reason for the relatively broad distribution of $Z^*$ for each element is that the DC-light intensity depends not only on the charge of the primary, but also on the emission height of the DC-light (Fig. \ref{nphot}). The mean emission height is determined by the first interaction height distribution of the primary particles in the atmosphere. As mentioned this varies significantly from event to event, with a FWHM of $\approx$10 km. This dependence on the emission height is one of the reasons for the observed bias in the charge reconstruction at lower atomic numbers, since the mean first interaction height varies between elements. A second reason is that the reconstruction of $Z^*$ is normalized to the charge of iron, which has not saturated its DC-emission in this energy region. 

In principle a more accurate measurement of the primary charge could be achieved by constraining the mean emission height $h_{\rm{DC}}$. This is possible because $h_{\rm{DC}}$ is directly related to $\Delta_{\rm{DC}}^{\rm{dir}}$ and $R_{\rm{core}}$ via:
\begin{equation}
h_{\rm{DC}} \simeq \frac{R_{\rm{core}}}{\Delta_{\rm{DC}}^{\rm{dir}}} ,
\end{equation}
This equation follows directly from the geometry of the emission, where $\Delta_{\rm{DC}}^{\rm{dir}}$ is the mean Cherenkov angle under which the DC-light is emitted.  However, in the case of the H.E.S.S. telescope system, little is gained by including the $h_{\rm{DC}}$ dependence in $Z^*$, since the pixel size of 0.16$^\circ$ and the spread of the direction and impact parameter reconstruction for DC-events are too coarse to provide a sufficiently precise determination of $\Delta_{\rm{DC}}^{\rm{dir}}$ and $R_{\rm{core}}$. It should be noted that the limited reconstruction accuracy of $R_{\rm{core}}$ also limits other techniques, such as that proposed in \cite{KIEDA}, which rely on this parameter.

\section{Systematic checks}

\begin{figure*}
  \begin{center}
  \resizebox{0.48\hsize}{!}{\includegraphics{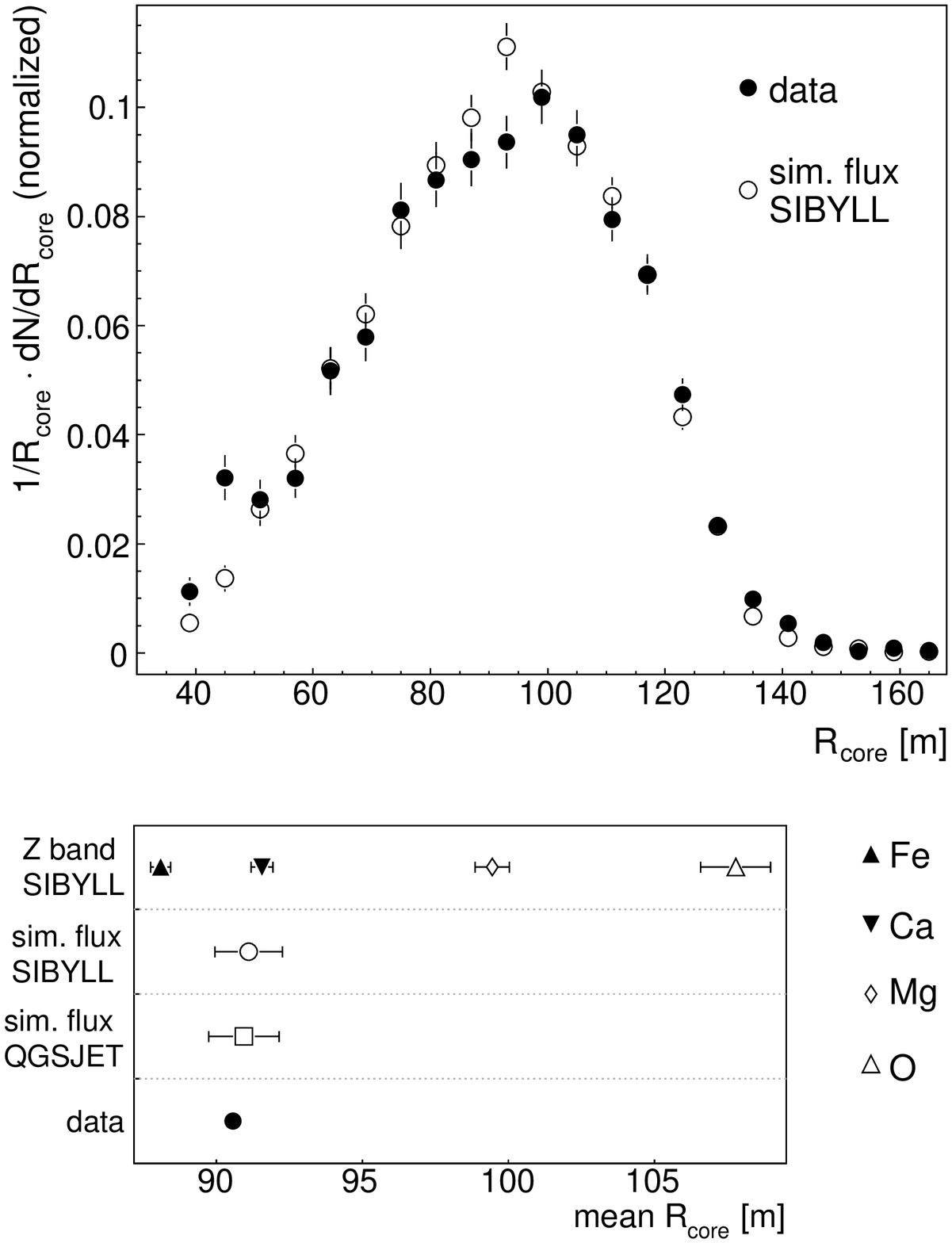}} 
  \resizebox{0.48\hsize}{!}{\includegraphics{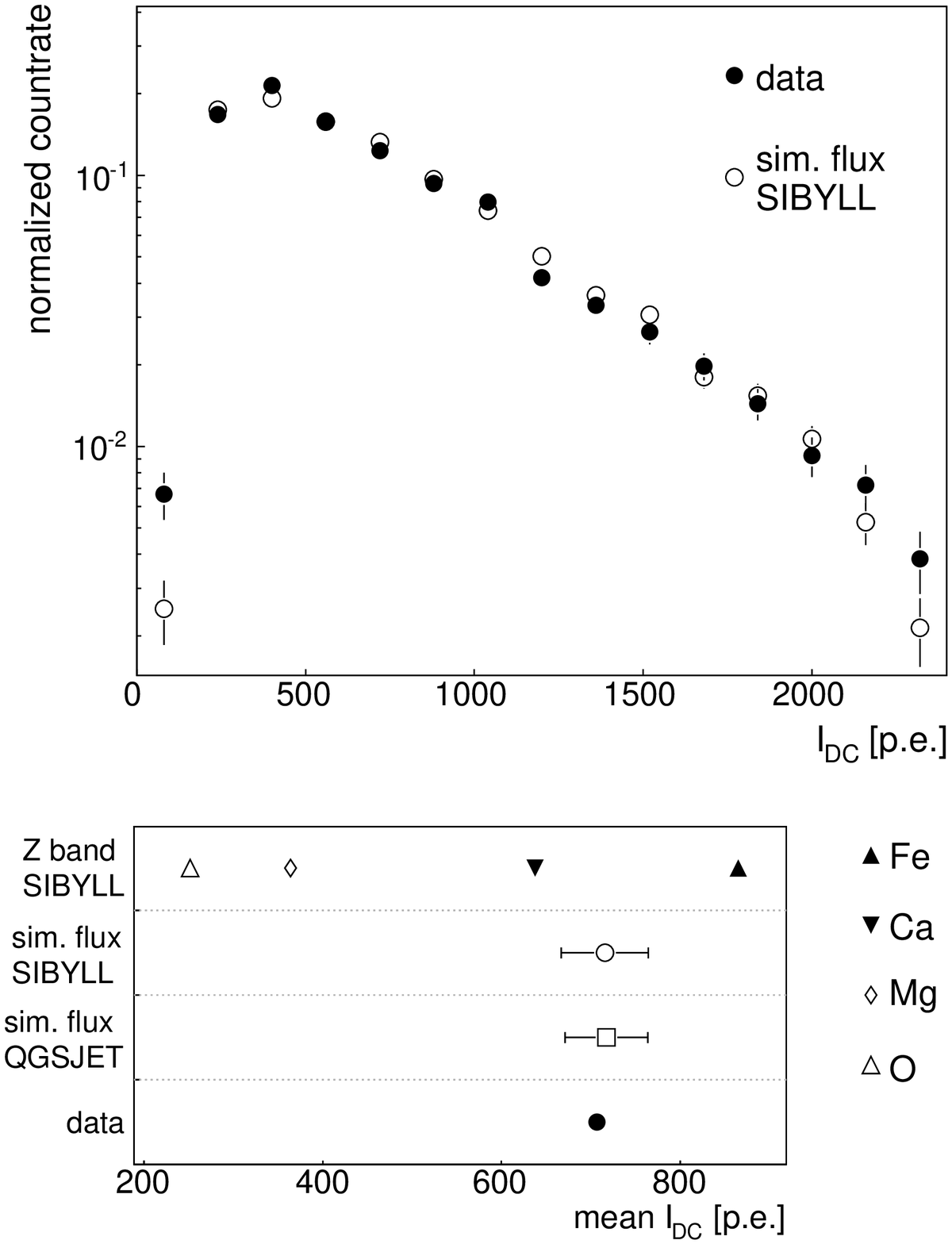}}
  \resizebox{0.48\hsize}{!}{\includegraphics{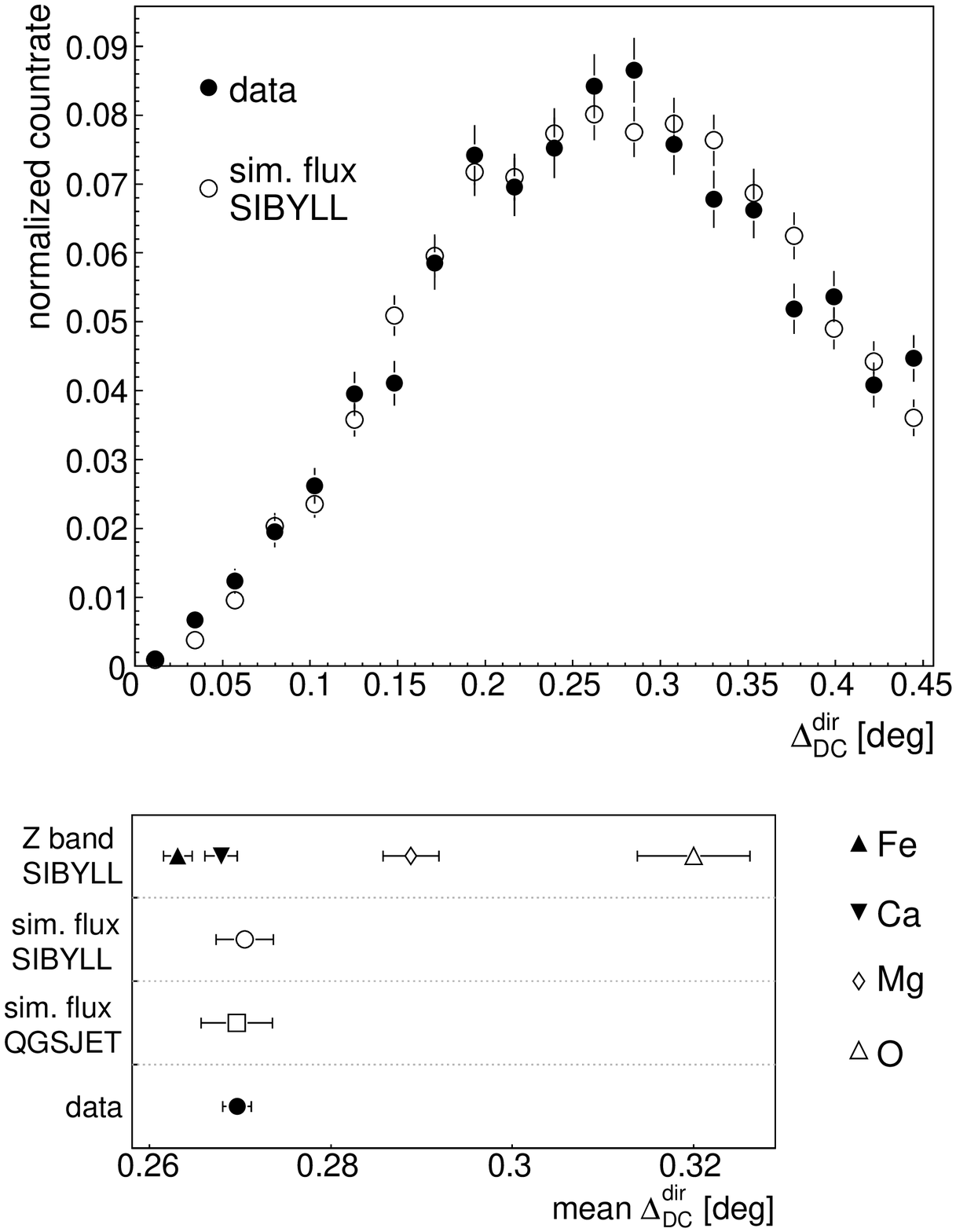}}
  \resizebox{0.48\hsize}{!}{\includegraphics{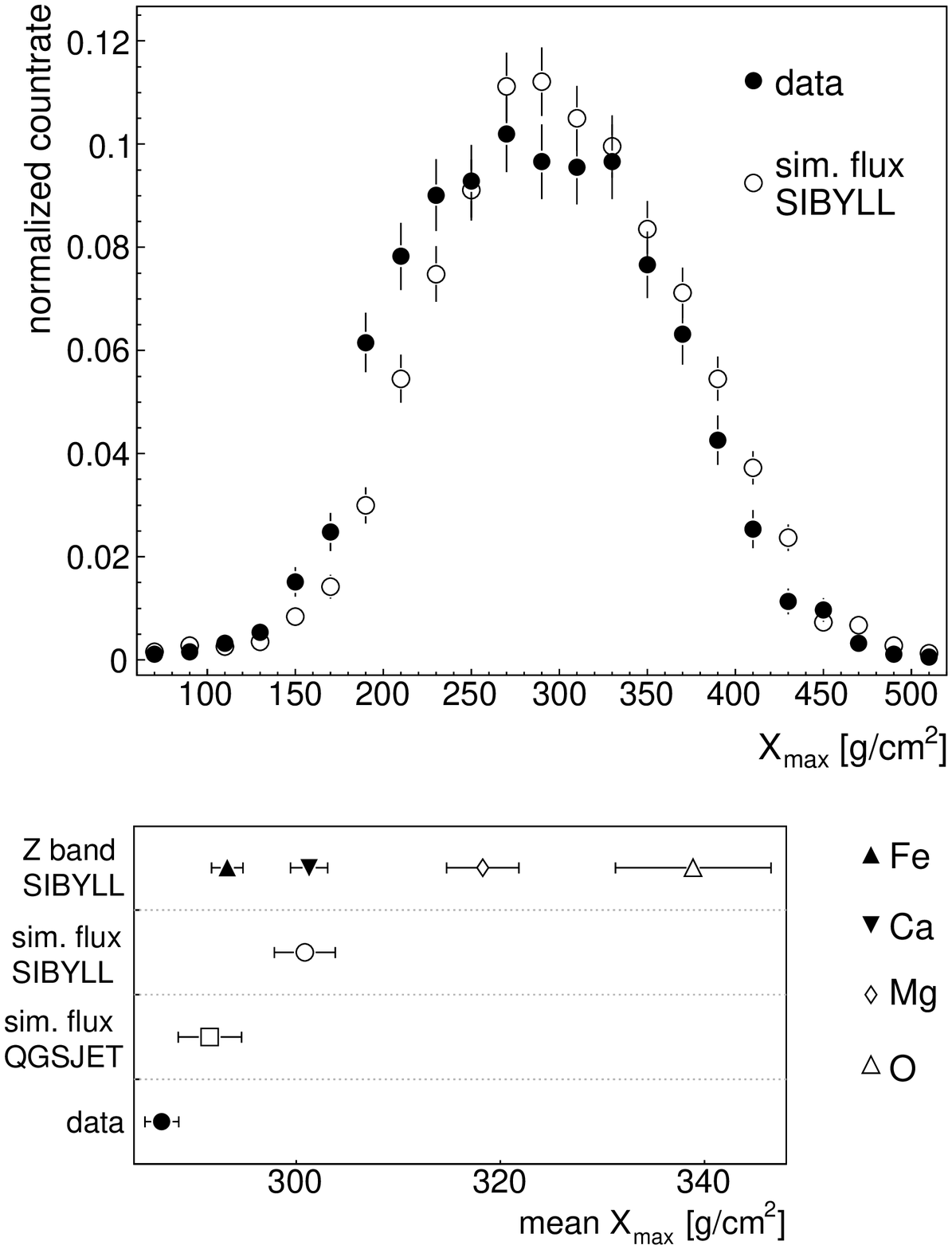}}
  \end{center}
  \caption{ Comparison of shower parameter distributions for data and simulations of the cosmic-ray flux. The bigger panels show normalized distributions of $R_{\rm{core}}$, $I_{\rm{DC}}$, $\Delta_{\rm{DC}}^{\rm{dir}}$ and $X_{\rm{max}}$ for SIBYLL simulations and data. The smaller panels underneath the distributions  show their mean values and the mean value for the different charge bands of which the simulated flux is  composed.  Additionally, they show the mean value for the distributions for the QGSJET simulations. When comparing the mean values for the simulated fluxes one should bear in mind that roughly equal contributions to the error bars come from statistical errors and from uncertainties in the reference composition, which is the same for both models. The systematic difference between the mean values for the shower maximum $X_{\rm{max}}$ is not unexpected, since this quantity is difficult to treat in the simulations (see text).}
  \label{param}
\end{figure*}

Since the reconstruction of the energy spectrum of cosmic-ray iron relies on Monte Carlo simulations, it is important to demonstrate adequate agreement between the measured and simulated distributions of parameters used in the reconstruction. Fig. \ref{param} shows such comparisons for the impact parameter, the DC-light intensity, the mean Cherenkov angle and the shower maximum $X_{\rm{max}}$. The shower maximum is the atmospheric depth at which the maximum number of Cherenkov photons is emitted in the EAS. The panels beneath the distributions show the mean values of the distributions for the data and both hadronic models. The error bars on the simulated points include both the statistical uncertainties and the uncertainty in the cosmic-ray mass composition in this energy regime.
 
The distributions of $R_{\rm{core}}$, $I_{\rm{DC}}$ and $\Delta_{\rm{DC}}^{\rm{dir}}$ show a good agreement between data and simulations. Their mean values agree within 1 $\sigma$. For the height of the shower maximum $X_{\rm{max}}$,  a shift of $\approx$5\% between SIBYLL simulations and the data is apparent. However, no significant shift in this parameter is present in the QGSJET simulations. This difference between the models is again an estimate for the systematic error arising from hadronic interaction uncertainties. Within this systematic uncertainty data and simulations agree reasonably well for $X_{\rm{max}}$. However, that the mean $X_{\rm{max}}$ values for both hadronic models are larger than the data is an indication that there might be additional systematic uncertainties in the reconstruction of this parameter. These could for example come from the uncertainty in the atmospheric profile at the H.E.S.S. site, which is estimated to be $\sim$3 g cm$^{-2}$. However, since no cut is applied on $X_{\rm{max}}$, these systematic effect could only affect the results indirectly and must be smaller than the hadronic model uncertainties. 

The larger difference between the two hadronic models for $X_{\rm{max}}$ compared to the other parameters shown is not completely unexpected. The shower maximum depends on the exact modeling of the fragmentation processes in the EAS-shower. The differences in the fragmentation process between hadronic models are well known (see for example \cite{MODELS}, \cite{JIM}). In contrast to the shower maximum, the distributions of $R_{\rm{core}}$, $\Delta_{\rm{DC}}^{\rm{dir}}$ and $I_{\rm{DC}}$ of the DC-events are dominated by the properties of the DC-light.  These are easier to model because they are completely determined by the distribution of the first interaction height for a given energy and atmospheric profile. Since composition measurements with the DC-light technique rely primarily on the DC-light intensity, they are expected to be relatively model independent. 

\section{Spectrum extraction}
\subsection{Iron fraction}

\begin{figure}
  \begin{center}
    \includegraphics[width=8cm]{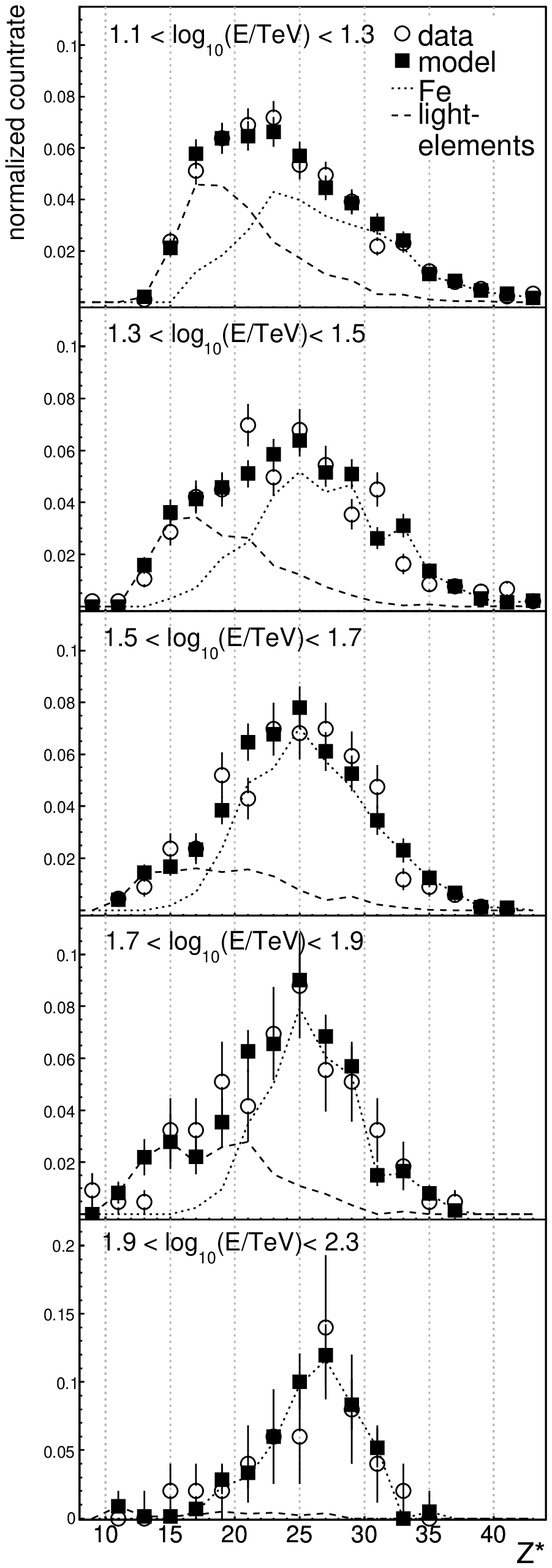}
  \end{center}
  \caption{$Z^*$ distributions of data (open circles) and the fitted SIBYLL model (black boxes) in five energy bands. The curves show the charge distributions of iron (dotted) and of the lighter charge bands (dashed) that compose the fitted model. The fit results are summarized in Table \ref{fitresults}. Note that, as mentioned in section \ref{sec:charge}, the charge resolution improves with energy, observable by a narrowing of the $Z^*$ distribution of iron with increasing energy.}
  \label{figfits}
\end{figure}

\begin{table*}
\begin{tabular}{l|c|c|c|c}
&$\log_{10}$(E/TeV) & $k^{\rm{data}}_{\rm{Fe}}$  $\pm$  $\Delta k_{\rm{fit}}$ $\pm$  $\Delta k_{\rm{comp}}$ & $\chi^2_{\rm{data}}$ /ndf & $k^{\rm{ref}}_{\rm{Fe}}$\\ \hline
\multirow{5}{*}{\begin{sideways}\centering SIBYLL\end{sideways}}
&1.1-1.3&0.56$\pm$0.047$\pm$0.026&6.9/15&0.55$\pm$0.09 \\
&1.3-1.5&0.64$\pm$0.049$\pm$0.027&31.5/17&0.64$\pm$0.08 \\
&1.5-1.7&0.77$\pm$0.054$\pm$0.019&15.4/15&0.70$\pm$0.07 \\
&1.7-1.9&0.66$\pm$0.097$\pm$0.013&12.0/14&0.80$\pm$0.06 \\
&1.9-2.3&0.93+0.07-0.151$\pm$0.008&3.4/12&0.84$\pm$0.04 \\
\hline
\multirow{5}{*}{\begin{sideways}QGSJET\end{sideways}} 
&1.1-1.3&0.47$\pm$0.050$\pm$0.026&11.6/15&0.50$\pm$0.10 \\
&1.3-1.5&0.55$\pm$0.059$\pm$0.029&13.3/17&0.66$\pm$0.08 \\
&1.5-1.7&0.70$\pm$0.063$\pm$0.023&14.3/15&0.75$\pm$0.06 \\
&1.7-1.9&0.54$\pm$0.128$\pm$0.034&9.3/14&0.79$\pm$0.06 \\
&1.9-2.3&0.69$\pm$0.160$\pm$0.012&6.2/12&0.84$\pm$0.05 
\end{tabular}
\caption{The best fit value of the iron fraction in the data $k_{\rm{Fe}}$ and the $\chi^2$ values of the fit are shown for both hadronic models in five energy bands. The error of $k_{\rm{Fe}}$ is composed of the statistical error of the fit $\Delta k_{\rm{fit}}$ and the systematic error from the uncertainty in the assumed composition of the lighter nuclei $\Delta k_{\rm{comp}}$ (see text). Additionally shown for comparison is the fraction of iron in the simulated cosmic-ray flux ($k^{\rm{ref}}_{\rm{Fe}}$) for both models.}
\label{fitresults}
\end{table*}

The first step in the derivation of the flux of iron nuclei is the measurement of fraction $k_{\rm{Fe}}$ of iron events among the DC-events. $k_{\rm{Fe}}$ is estimated via a fit of a two-component model to the $Z^*$ distribution of the data. The first component of this model is the $Z^*$ distribution of simulated iron nuclei. The second component is a sum of the $Z^*$ distribution of lighter nuclei. The relative composition of the lighter charge bands (= all except the iron band) is kept fixed to the reference composition, so that $k_{\rm{Fe}}$ is the single free parameter of the fit.

The fit was performed in five energy bands. The $Z^*$ distribution from the data and the fitted model using SIBYLL simulations are shown in Fig. \ref{figfits}. The fit results ($k_{\rm{Fe}}$) and corresponding $\chi^2$ values ($\chi^2_{\rm{data}}$) for the fits are summarized in the Table \ref{fitresults} for QGSJET and SIBYLL. The values for $k_{\rm{Fe}}$ for the two hadronic models agree with each other within statistical errors $\Delta k_{\rm{fit}}$ (the standard deviation of the fit result) for individual energy bands. However, the QGSJET values are shifted by $\approx$-0.1.

Apart from statistical uncertainties,  $k_{\rm{Fe}}$ is affected by the systematic uncertainty in the assumed composition of the lighter nuclei $\Delta k_{\rm{comp}}$. This is estimated by varying the weight of the individual lighter charge bands of the fitted model by 25\% and performing the fit for each possible combination. The minimum and maximum deviation between these fit results and the previously obtained $k_{\rm{Fe}}$ value are then taken as errors on the composition uncertainty. Since these errors are close to symmetric, their absolute values are averaged to give $\Delta k_{\rm{comp}}$.

Table \ref{fitresults} also shows the expected iron fractions $k_{\rm{Fe}}^{\rm{ref}}$ from simulations of the cosmic-ray flux assuming the  reference composition. These values agree with the best fit values. This implies, together with the reasonable $\chi^2 / ndf$ values of the fits, that no significant deviation from the reference composition can be found in the data. For the iron band this statement will be quantified in the next section.       

\subsection{Iron Flux}
\label{sec:flux}

\begin{figure*}
  \begin{center}
    \includegraphics[width=18cm]{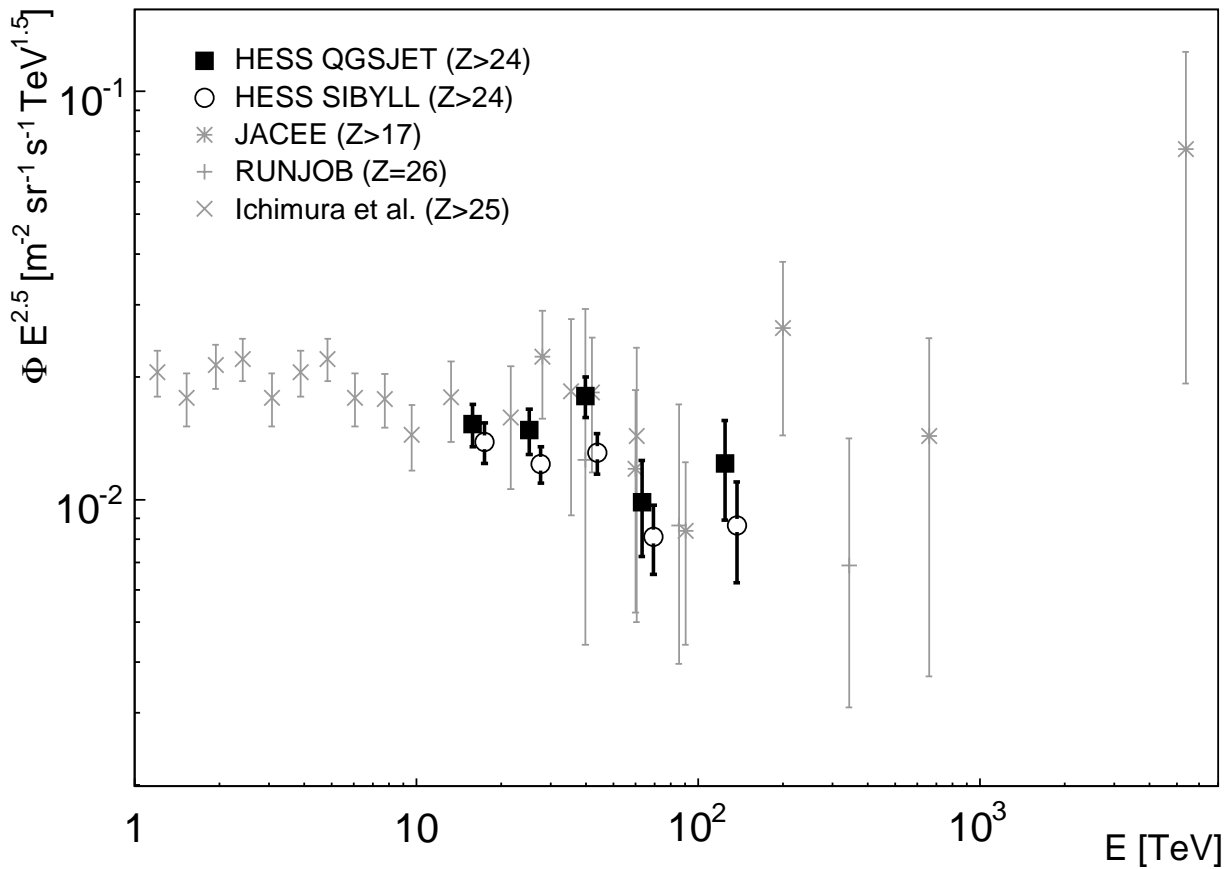}
  \end{center}
  \caption{Differential iron energy spectrum measured with H.E.S.S. for the hadronic models QGSJET and SIBYLL multiplied by E$^{2.5}$ for better visibility of structures. The  spectral points for both models are measured for the same energies. For better visibility  the SIBYLL points were shifted 10\% upwards in energy. The error bars show the statistical errors.   The systematic flux error in each bin is 20\%. The measurements from balloon experiments with data points at the highest energies are shown for comparison \cite{JACEE,RUNJOB,ICHIMURA} (a compilation with more measurements from balloon experiments and space born measurements can be found in \cite{HOERANDEL}). For a better visibility no horizontal bars marking the bin ranges are shown, they can be found in the respective papers. When comparing the measurements one should bear in mind that the experiments have different charge thresholds for their definition of the iron band (see legend). }
  \label{figspectrum}
\end{figure*}

Since the identification of DC-events is effectively background-free, the differential iron flux $\phi(E)$ can be estimated as:
\begin{equation}
\phi(E)  = \frac{N_{\rm{DC}}(E)}{A_{\rm{eff}}(E)\cdot \Delta E \cdot  t}\cdot k_{\rm{Fe}} ,
\end{equation}
where $N_{\rm{DC}}(E)$ is the number of detected DC-events in the energy interval from $E$ to $E+\Delta E$, $t$ is the total live-time of the data-set and  $A_{\rm{eff}}$ is the mean effective area times the field of view of the detector, averaged over the zenith angle of the observations, taking into account the efficiency of selection cuts. $A_{\rm{eff}}$ is derived from simulations of iron nuclei via:
\begin{equation}
A_{\rm{eff}}  = \frac{ N^{\rm{MC}}_{\rm{DC}}(E) \cdot A^{\rm{MC}} \cdot \Omega^{\rm{MC}}}{N^{\rm{MC}}(E)} ,
\end{equation}
where $N^{\rm{MC}}(E)$ is the total number of simulated events in the energy interval from $E$ to $E+\Delta E$ and $N^{\rm{MC}}_{\rm{DC}}$ is the corresponding number of identified DC-events. $A^{\rm{MC}}$ and $\Omega^{\rm{MC}}$ are the area and angular region over which the simulations were performed. 

The energy spectrum is measured in the five energy bins of the $k_{\rm{Fe}}$ fit. The result is shown in Fig. \ref{figspectrum} for both hadronic models together with the highest energy baloon measurements.  The derived spectrum agrees well with these measurements for both models. The measured spectrum is fitted well by a power law $\phi(E) =\phi_0 (\frac{E}{\rm{TeV}})^{-\gamma}$. The best fit values for the SIBYLL spectrum are given by
$\phi_0 = (0.029 \pm 0.011)$ m$^{-2}$sr$^{-1}$ TeV$^{-1}$ and 
$\gamma = 2.76\pm0.11$  with an 
$\chi^2$/ndf of 3.0/3. For the QGSJET spectrum the best fit values are 
$\phi_0 = (0.022 \pm 0.009)$ m$^{-2}$sr$^{-1}$ TeV$^{-1}$ and 
$\gamma = 2.62 \pm 0.11 $  with 
$\chi^2$/ndf of 5.3/3. The integrated flux above 13 TeV is 
$F(>13 \rm{TeV}) = (1.9 \pm 0.7) \cdot 10^{-4}$ s$^{-1}$sr$^{-1}$ m$^{-1}$ for SIBYLL and 
$F(>13 \rm{TeV}) = (2.3 \pm 0.9) \cdot 10^{-4}$ s$^{-1}$sr$^{-1}$m$^{-1}$ for QGSJET.  

Since both spectra are derived using the same data-set, the differences in the spectral index $\Delta \gamma = 0.14$ and integrated flux $\Delta F/F = 17 \%$ again provide an estimate of the systematic error due to hadronic interaction uncertainties. Additional systematic errors, arising from uncertainties in the atmospheric profile and the absolute detection efficiency of the H.E.S.S. instrument are discussed in detail in \cite{CRAB} and lead to a  systematic error of 20\% in the integrated flux and $\Delta \gamma = 0.1$ in the spectral index. The effect of the systematic error $\Delta k_{\rm{comp}}$ in $k_{\rm{Fe}}$ on the spectrum amounts to  $\Delta \gamma = 0.015$ and $\Delta F/F = 5\%$. Assuming a more conservative error of 50\% in the integral fluxes of the lighter elements in the model of the $k_{\rm{Fe}}$ fit increases $\Delta k_{\rm{comp}}$ by 0.03 on average and leads to  errors of  $\Delta \gamma = 0.04$ and $\Delta F/F = 11\%$ in the presented spectrum. This error is still small compared to the previously mentioned uncertainties.  The total systematic uncertainty of the measurement is therefore estimated to $\Delta \gamma = 0.17$ for the spectral index and $\Delta F/F = 28\%$ for the integrated flux.

The statistical error on the measured iron flux is comparable to these systematic errors. This means that without an improvement in the latter,  the total error of the measurement can not be significantly reduced by increasing the exposure time of the data set. However, an increased data-set would enable one to extend the measurement towards higher energies. We note that, despite the systematic uncertainties, the iron flux determined with this technique is one of the best measurements in this energy range. The good agreement between the measured fluxes from balloon experiments and those given here lends confidence to the results from both techniques.

\section{Summary \& Outlook}

A technique for the detection of cosmic rays by resolving the Cherenkov emission from primary particles  has been presented and applied to H.E.S.S. data. As a result 1899  events  with Direct Cherenkov light in at least two telescopes were detected and it was shown that these DC-events can be considered as background-free. Different parameter distributions of these events were compared to simulations using two different hadronic interaction models and good agreement with the data was found for both. The strong correlation between the DC-light and the charge of the primary shower particle made a charge estimate possible, from which the energy dependent fraction of iron in the data was derived. The energy spectrum of iron nuclei was determined in an energy range of 13 to 200 TeV. The result confirms the flux measurements from balloon experiments with an independent technique and is one of the most precise measurements in this energy range.

Future improvements of the DC-light technique could extend the energy range of the measurement to an energy of $\sim 1$ PeV. Besides larger statistics, this extension requires additional separation power of the DC-light  from the EAS-light. The reason for this is that the DC-light yield remains constant above a certain energy while the EAS light yield increases approximately linearly with energy. As shown in \cite{KIEDA}, additional separation power can be achieved using the time structure of the DC-light, since it arrives with a typical delay of 4 ns with respect to the EAS light. This fact could not be exploited in the analysis presented  because the H.E.S.S. data used here were taken with the standard integration window of 16 ns. However, current and planned Cherenkov telescopes, which routinely store pulse timing information \cite{MAGIC,VERITAS}, may take advantage of this characteristic. 

Due to the strong dependence of the DC-light yield on the charge of the primary particle, the DC-light technique has great potential for composition measurements. The limiting factor is currently the accuracy of shower reconstruction, needed to constrain the emission height of the DC-light. Since the typical shower images in the present work contain $\sim$100 pixels, the limitation in the reconstruction accuracy of the shower with the Hillas technique  arises from the strong fluctuations in hadronic showers and not from the limited angular resolution of the system. This puts a physical limit to the charge resolution when the showers are reconstructed using this technique.

In order to reduce the sensitivity to shower fluctuations, the DC-light itself could in principle be used to reconstruct the shower. One simple extension of the current method would be to determine the principal axes of the shower images as the line connecting the center of gravity of the image to the DC-pixel. However, exploitation of this technique would require pixels with smaller angular scale for the accurate localization of the DC-light spot.  Another possibility, with an array of many nearby telescopes, would be to reconstruct the shower from the DC-light intensity distribution on the ground. Both techniques would take advantage of the very small fluctuations of the DC-emission. A quantitative statement on the level of improvement for both techniques would require detailed simulations and is beyond the scope of this paper.

Finally, the agreement of the distribution of the reconstructed charge $Z^*$ for large charges ($Z^*> 28$) disfavors a significant contribution from ultra heavy elements to the cosmic-ray flux, as expected. It should also be noted that the present analysis is not sensitive to exotic states of matter, such as \emph{quark matter} or \emph{magnetic monopoles} as proposed in \cite{UHCR}. The high charges of these states ($Z \gg 100$) imply DC-intensities which would saturate the photomultipliers in the H.E.S.S. cameras.

\begin{acknowledgments}
The support of the Namibian authorities and of the University of Namibia
in facilitating the construction and operation of H.E.S.S. is gratefully
acknowledged, as is the support by the German Ministry for Education and
Research (BMBF), the Max Planck Society, the French Ministry for Research,
the CNRS-IN2P3 and the Astroparticle Interdisciplinary Programme of the
CNRS, the U.K. Particle Physics and Astronomy Research Council (PPARC),
the IPNP of the Charles University, the South African Department of
Science and Technology and National Research Foundation, and by the
University of Namibia. We appreciate the excellent work of the technical
support staff in Berlin, Durham, Hamburg, Heidelberg, Palaiseau, Paris,
Saclay, and in Namibia in the construction and operation of the
equipment. We would also like to thank Joerg Hoerandel for providing data on cosmic-ray fluxes in numerical form.
\end{acknowledgments}

\end{document}